# Ferroelectricity in Single Crystal InMnO$_3$

T. Yu[1], P. Gao[1], T. Wu[1], T. A. Tyson[1], and R. Lalancette[2]

[1]Department of Physics, New Jersey Institute of Technology, Newark, New Jersey 07102

[2]Department of Chemistry, Rutgers University, Newark, New Jersey 07102

Corresponding Author: T. A Tyson, E-mail: tyson@adm.njit.edu

## Abstract

Single crystal synthesis, structure, electric polarization and heat capacity measurements on hexagonal InMnO$_3$ show that this small R ion in the RMnO$_3$ series is ferroelectric (space group P6$_3$cm). Structural analysis of this system reveals a high degree of order within the MnO$_5$ polyhedra but significant distortions in the R-O bond distributions compared to the previously studied materials. Point-charge estimates of the electric polarization yield an electrical polarization of ~7.8 $\mu$C/cm$^2$, 26% larger than the well-studied YMnO$_3$ system. This system with enhanced room temperature polarization values may serve as a possible replacement for YMnO$_3$ in device application.

The hexagonal manganites are part of a class of materials which exhibit coupled ferromagnetism and ferroelectricity (multiferroic systems).[1] Hexagonal phase $RMnO_3$ is found for small radius ions (R=Ho, Er, Tm, Yb, Lu and also Y and Sc, see Refs. 2, 3, 4). From an application's perspective, these materials have attracted much attention as data storage media in nonvolatile random access memory.[5] These devices have the advantage of low power consumption and decreased memory cell size over existing technology because the electric charge induced by the remnant polarization controls the conductivity of the Si substrate where they are deposited. Low dielectric constant materials such as the hexagonal $RMnO_3$ systems ($YMnO_3$, $\varepsilon = 20$) do not suffer from the problem of high electrical noise due to $SiO_2$ at the interface. The $YMnO_3$ system has been studied extensively and is currently being utilized in device applications.[6]

Tuning this $RMnO_3$ system by replacing the R ion with other species or multiple species serves as a possible path to enhance the ferroelectric properties of these materials. One system which is being studied is the R = In system. However, the room temperature structure and ferroelectric properties of $InMnO_3$ are still not well understood. $InMnO_3$ x-ray single crystal structural refinement by Giaquinta *et al.*[7] yielded a non-polar space group of $P6_3/mmc$ for samples prepared by the flux method similar to earlier work on single crystals. The group also reported neutron diffraction measurements for polycrystalline samples prepared by a nitrate decomposition technique and found a space group $P6_3cm$. Fabrèges *et al.*[8] reported the space group $P6_3cm$ based on neutron powder diffraction measurements on polycrystalline samples prepared by high pressure solid state synthesis. Kumagai *et al.*[9] prepared polycrystalline samples of $InMnO_3$ by solid state synthesis and studied them by synchrotron x-ray powder diffraction methods. A significant impurity phase of $In_2O_3$ was observed. Equally good fits to both the polar P6$_3$cm and non-polar P-3c structure were found. In this work, second harmonic generation techniques found no evidence for domains significantly above the micron length scale. DFT simulations indicated that the P-3c structure has a lower energy than the P6$_3$cm structure (lower by 200 meV per formula unit) and the combined results led the authors to conclude that P-3c symmetry was appropriate to this system at room temperature. Previous electrical measurements were conducted on polycrystalline materials and showed strong leakage behavior but no clear signal of ferroelectricity.[10,11]

To fully understand the properties of the InMnO$_3$ system, single crystal samples were grown and the structure as well as the thermal and electronic properties was studied. The space group of single crystal material is found to be consistent with the previous neutron powder diffraction measurements and a material with a P6$_3$cm polar space group [Fig. 1(c)] is revealed. Magnetic ordering temperatures are identified by heat capacity measurements. Polarization measurements show a finite value of polarization for the first time. Point charge and density functional theory calculations of the electric polarization reveal an upper limit of 7.8 $\mu$C/cm$^2$ on the electrical polarization, which is 26% larger than that for YMnO$_3$.

InMnO$_3$ single crystals were grown by the flux method starting from In$_2$O$_3$ and MnO$_2$ in a 1:2 molar ratio, with 1 part Bi$_2$O$_3$ added relative to In$_2$O$_3$. After the oxides were ground together and pelletized, the pellets were calcined at 950 °C for 3 days in air in a platinum crucible and quenched in to ambient temperature by taking the crucible out of the furnace at 950 °C. Black hexagonal plates [Fig. 1(a)] were visible on the surface of the pellet. The flux matrix was mechanically weakened and dissolved with concentrated nitric acid by soaking for 2 weeks and cleaned in an ultrasonicator with acetone.[7(b)] The cleaned single crystals were examined by EDX and XAFS and surface traces of Bi were found but verified as amorphous by x-ray absorption spectroscopy. Single crystal x-ray diffraction measurements were conducted utilizing a Bruker APEX II diffractometer with 4K CCD detector and Cu K$_\alpha$ radiation (see details below). A single crystal of dimensions ~0.214 × 0.263 × 0.025 mm$^3$ was used [Fig. 1(b)].

Polycrystalline samples of hexagonal RMnO$_3$ (R=Sc, Y, and Lu) were prepared by the solid state reaction method. InMnO$_3$ power was obtained by grinding single crystals. X-ray absorption samples were prepared by brushing powder (500 mesh) onto Kapton tape and spectra were collected at beamline X19A, the National Synchrotron Light Source. A Mn foil reference was employed for energy calibration. The reduction of the x-ray absorption fine-structure (XAFS) data was performed using standard procedures.[12]

To estimate the electrical polarization, the reference structure at high temperature was determined by full structural optimization with density functional calculations (DFT) in the projector augment wave approach[13] holding the Mn and In ions at high symmetry positions (Mn at (1/3,0,0), In1 at (0,0,1/4), and In2 at (1/3,2/3,1/4)) following the methods used in Ref. 14.

Electrical polarization loops of an InMnO$_3$ single crystal (thickness ~25 $\mu$m, sample area 0.26 × 0.54 mm$^2$, and small electrode area ~8991 $\mu$m$^2$) were obtained using a Radiant Technology Multiferroic instrument with the remanent hysteresis method at room temperature. Silver paste was used to connect to the electrodes on both sides. We achieved similar results from other samples in the thickness range of 15 $\mu$m to 30 $\mu$m for frequencies 500 to 2000 Hz. Heat capacity measurements of an InMnO$_3$ single crystal from 300 K to 2 K were carried out with a Physical Property Measurement System (PPMS, Quantum Design).

Fig. 2(a) shows the near edge x-ray absorption spectra of a series of hexagonal RMnO$_3$ systems (R=Sc, Y and Lu) compared to InMnO$_3$. Note the similarity in shape of all of the spectra indicating equivalent local structure and local symmetry. Compared to the other samples, no additional peaks appear in the R=In spectrum. However, the InMnO$_3$ spectrum has lower amplitude and is broadened. This indicates a lower level of long-range order. Note the shift in the position of the main peak of the ScMnO$_3$ to higher energy compared to the other samples. This shift shows that the Mn-O bond distance is shorter in this system than the others according to the "Natoli's Rule", $(E_p-E_0)R^2$=const., where $E_p-E_0$ is the energy of the peak measured from $E_0$, and $R$ is bond distance.[15]

The Fourier transform of the full XAFS data is shown in Fig. 2(b). Note that the peak for Mn-O bond is the largest for ScMnO$_3$ and InMnO$_3$ indicating high structural order within the MnO$_5$ polyhedra. The second peak (R ~3 Å) corresponds to Mn-R and Mn-Mn (first neighbor) and depends on the R ions' scattering power. The distant peak (R ~6.5 Å) is dominated by Mn-Mn correlations and is significantly suppressed in the case InMnO$_3$, indicating weak higher order Mn-Mn correlation. More structural details can be gained from the analysis of the single crystal x-ray diffraction data.

Single crystal structure solution was conducted using SHELXL[16] after the data were corrected for absorption by face indexing.[17] Refinements with respect to P-3c, P3c1 and P6$_3$cm space groups were conducted. Use of the non-polar space group P-3c yielded R$_1$ = 6.12% and wR$_2$ = 15.6% with ratio of number of F$_0$ > 4 σ(F$_0$) values to free parameters of 199/21 = 9.5 while the polar space group P3c1 yielded R$_1$ = 4.81% and wR$_2$ = 12.8% and ratio of number of F$_0$ > 4 σ(F$_0$) values to free parameters of 371/40 = 9.3. Analysis of the coordinates revealed that this space group solution (P3c1) corresponded to the higher space group P6$_3$cm (Ref. 18 ). Refinement of the diffraction data in the polar space group P6$_3$cm yielded R$_1$ = 5.07% and wR$_2$ = 12.6% with ratio of number of F$_0$ > 4 σ(F$_0$) values to free parameters of 230/29 = 7.9. For this latter structure, racemic twinning was modeled by SHELXL yielding weights of 0.44 and 0.56 for the twin components (structure and its inverse). This is close to the 50% value expected for ferroelectric samples. Use of Cu Kα radiation produced non-negligible anomalous scattering factors[19] from the In (f′=0.0822, f″=5.0449) and Mn (f′=-0.5299, f″=2.8052) ions making the Friedel pairs distinct thus stabilizing the refinement with respect to the twin structure. A Bayesian statistical analysis[20] of the Friedel (Bijvoet) reflection pairs revealed a 99.9% probability of the 1:1 racemic twin model being correct and negligible probability for either component of the pair being the correct solution. We note that racemic twinning was also found in the $YMnO_3$ and $YbMnO_3$ ferroelectric systems with $P6_3cm$ space group.[21] The existence of this twinning is expected for a ferroelectric system with no bulk polarization (not poled). Details for the structural solution are given in Table I, and the bonding compared to the systems with R=Y, Sc and Lu[22] are given in Table II. Compared to $ScMnO_3$ [<Mn-O> = 1.932(3) Å], $YMnO_3$ [<Mn-O> = 1.984(14) Å], and $LuMnO_3$ [<Mn-O> = 1.966(8) Å], the coordination of Mn by O revealed by the bond distance is closest to $ScMnO_3$, consistent with the XAFS results. In addition the distribution of Mn-O bonds ($\varepsilon_{Mn-O}$, Table II) is significantly more narrow (more ordered) for $InMnO_3$ than any of these other systems. We also note that the first neighbor Mn-Mn distribution ($\varepsilon_{Mn-Mn}$, Table II) is the smallest for $InMnO_3$. On the other hand, the distribution of R-O is much larger for R=In than for the other systems. Since the coordination of the R site which primarily drives the electrical polarization, it suggests a much higher polarization for $InMnO_3$. The structural data reveal that the system is polar with the same space group $P6_3cm$ of R=Sc, Y and Lu systems. To further confirm the assignment, electrical polarization measurements were conducted.

Polarization hysteresis loops were calculated by integrating the total transferred charge during application of a bipolar triangular voltage waveform, then dividing the remanent charge by the projected area as done in Ref. 23. We examined a large range of electric fields from 64 kV/cm to 216 kV/cm and displaying the P-E loops, which give a remanent polarization (Pr) of ~4.4 $\mu$C/cm$^2$ for the saturated loop at room temperature shown in Fig. 3. The remanent polarization exhibits negligible frequency dependence in the range 500 Hz to 2000 Hz. The polarization was estimated by the point-charge model with the experimental structure derived above and using DFT to estimate the reference paraelectric structure. The polarization amplitude of 7.8 $\mu$C/cm$^2$ on InMnO$_3$ single crystal was obtained indicating a theoretical upper limit ~26% larger than that for YMnO$_3$ (near 6.2 $\mu$C/cm$^2$)[24]. Our measurement gave a value of ~4.4 $\mu$C/cm$^2$ for the remanent polarization at room temperature. The smaller value is possibly due to defect/voids formed during the quenching procedure. Alternative methods could be developed to prepare defect-free crystals or films for applications. We note that defects result in a range of values of the remanent polarization in YMnO$_3$ single crystals ~4.5 to 5.5 $\mu$C/cm$^2$.[5(d), 25]

Finally, for comparison with other systems, the heat capacity of InMnO$_3$ single crystals is shown in Fig.4(a) with the Néel temperature near 118 K identical with the result of Belik *et al.*[10] on powder samples. We have also found a peak near 42 K consistent with the spin rotation on Mn seen in HoMnO$_3$ (See Refs. 26). However, we note that nanoscale LuMnO$_3$ (See Ref. 27) with a closed 4*f* shell at the R site exhibits the same reorientation transition near 42 K and indicates the spin orientation may not be only driven by the Mn-R spin interactions as suggested by earlier studies[24]. The similarity of our study with the nanoscale measurement is that the nanomaterials are most likely single domain (structurally). The reduced lengthscale in these nanomaterials is similar with the lengthscale creating suppressed higher order Mn-Mn/R-O correlations seen in the XAFS and XRD measurements. (We note that in manganites an impurity phase $Mn_3O_4$ may occur with a magnetic transition near 41 K. However, we did not find evidence for this in synchrotron base powder diffraction measurements on the samples. No un-indexed peaks were observed.)

Low temperature fitting of heat capacity gave $\gamma = 2.87 \pm 0.09$ mJ mol$^{-1}$ K$^{-2}$, $\beta = 0.06 \pm 0.01$ mJ mol$^{-1}$ K$^{-4}$ and $\beta_5 = 0.0014 \pm 0.0002$ mJ mol$^{-1}$ K$^{-6}$ with fit by $C_p/T = \gamma + \beta T^2 + \beta_5 T^4$ in Fig.4(b). The terms correspond to electronic phonon, and magnetic/anharmonic contributions to the heat capacity, respectively.[29] We note that neglecting the third term produced only small changes in the Sommerfeld coefficient $\gamma$ given by $\pi^2 k_B^2 N(E_F)/3$, where $N(E_F)$ is the density of states at the Fermi level.[28] Values of $\gamma = 1.6$ or 3.6 mJ mol$^{-1}$ K$^{-2}$ (two samples) for YMnO$_3$ single crystals[29] and $\gamma = 1.0$ mJ mol$^{-1}$ K$^{-2}$ for LuMnO$_3$ single crystals[2] have been reported. The similar value of $\gamma$ (associated with the number free carriers) for InMnO$_3$ to the other well studied systems, suggests that the main determinant in the polarization is the R-O distribution.

Single crystal synthesis, heat capacity, structure and electric polarization measurements of hexagonal InMnO$_3$ reveal for the first time that this small R ion material has space group P6$_3$cm and is ferroelectric. Point-charge estimates of electric polarization based on the single crystal derived structure predict an electrical polarization of 7.8 $\mu$C/cm$^2$, 26% larger the standard RMnO$_3$ systems. This system with enhanced room temperature polarization values may serve as possible replacement for YMnO$_3$ in device applications and, due to its interesting magnetic ordering, it may serve as a test-bed for theoretical studies of magnetic ordering.

This work is supported by DOE Grant DE-FG02-07ER46402. The Physical Properties Measurements System was acquired under NSF MRI Grant DMR-0923032 (ARRA award). X-ray diffraction and x-ray absorption data acquisition were performed at Brookhaven National

Laboratory's National Synchrotron Light Source (NSLS) which is funded by the U. S. Department of Energy.

**Table I. Structural Parameters from Single Crystal Refinement***

| Atoms | x | y | z |
|---|---|---|---|
| In1 | 0 | 0 | 0.2623(4) |
| In2 | 1/3 | 2/3 | 0.2328 |
| Mn | 0.3339(3) | 0 | -0.0107(18) |
| O1 | 0.3206(12) | 0 | 0.1545(41) |
| O2 | 0.6355(19) | 0 | 0.3244(33) |
| O3 | 0 | 0 | -0.0182(87) |
| O4 | 1/3 | 2/3 | 0.0135(62) |

| | | | | | |
|---|---|---|---|---|---|
| $U_{ij}$ (In1) | 0.0180(16) | 0.0180(16) | 0.0096(28) | 0.0000 | 0.0090(8) |
| $U_{ij}$ (In2) | 0.0165(11) | 0.0165(11) | 0.0140(22) | 0.0000 | 0.0082(5) |
| $U_{ij}$ (Mn) | 0.0183(13) | 0.0156(15) | 0.0049(13) | -0.0014(54) | 0.0078(7) |
| $U_{iso}$ (O1) | 0.0206(51) | | | | |
| $U_{ij}$ (O2) | 0.021(5) | 0.041(8) | 0.004(53) | -0.0015(62) | 0.021(4) |
| $U_{iso}$ (O3) | 0.013(13) | | | | |
| $U_{ij}$ (O4) | 0.013(10) | 0.013(10) | 0.024(25) | 0.0000 | 0.016(12) |

Space Group: $P6_3cm$
a = 5.8905(2) Å, c = 11.4824(5) Å, Dx = 6.288 g/cm$^3$
Measurement Temperature: 296 K
Crystal Dimensions: 0.214 × 0.263 × 0.025 mm$^3$
wavelength: 1.54178 Å,
2θmax: 142.2°
BASF twin parameter: 0.44(10)
Hooft y-parameter: 0.37(5)
Absorption Coefficient: 122.62 mm$^{-1}$
EXTI extinction parameter: 0.03477
Number of Unique Observed Reflections $F_o > 4\sigma(F_o)$: 230
Number of fitting parameters: 29
Amplitude of Max Peak in Final Difference map: 2.90 e/ Å$^3$ (In2)
$R_1$ = 5.07 %, $wR_2$ = 12.6 %, Goodness of Fit = 1.19

* Atomic displacement parameters $U_{ij}$ are in the order: $U_{11}, U_{22}, U_{33}, U_{13}, U_{12}$, ($U_{23}$ = 0 for all atoms by symmetry). For O1 and O3, only $U_{iso}$ values obtained.

**Table II. Selected Bond Distances (Å)***

| Bond Type | InMnO$_3$ | YMnO$_3$ | ScMnO$_3$ | LuMnO$_3$ |
|---|---|---|---|---|
| Mn-O1 | 1.904(50) | 1.850(15) | 1.884(7) | 1.882(7) |
| Mn-O2 | 1.907(40) | 1.878(15) | 1.876(7) | 1.859(7) |
| Mn-O3 | 1.969(5) | 1.996(11) | 1.968(5) | 2.050(5) |
| Mn-O4 x 2 | 1.982(10) | 2.097(6) | 1.967(5) | 2.019(7) |
| <Mn-O> | 1.949(2) | 1.984(14) | 1.932(3) | 1.966(8) |
| $\varepsilon_{Mn-O}*10^2$ | 3.568 | 10.476 | 4.286 | 7.897 |
| | | | | |
| Mn-R1 | 3.265(17) | 3.293(14) | 3.174(8) | 3.264(3) |
| Mn-Mn x 4 | 3.398(11) | 3.622(14) | 3.377(11) | 3.475(11) |
| Mn-Mn x 2 | 3.407(2) | 3.420(14) | 3.350(11) | 3.509(7) |
| Mn-R2 x 2 | 3.416(17) | 3.345(10) | 3.232(6) | 3.305(4) |
| Mn-R2 x 2 | 3.539(17) | 3.731(10) | 3.583(6) | 3.654(4) |
| Mn-R1 | 3.701(18) | 3.647(15) | 3.627(8) | 3.721(3) |
| <Mn-Mn> | 3.401 | 3.555 | 3.368 | 3.486 |
| $\varepsilon_{Mn-Mn}*10^2$ | 0.424 | 9.522 | 1.273 | 1.603 |
| | | | | |
| R1-O1 x 3 | 2.254(26) | 2.264(13) | 2.126(6) | 2.234(11) |
| R1-O2 x 3 | 2.261(13) | 2.353(9) | 2.223(3) | 2.294(14) |
| R1-O3 | 2.52(10) | 2.344(25) | 2.157(12) | 2.244(14) |
| <R1-O> | 2.295 | 2.314 | 2.172 | 2.261 |
| $\varepsilon_{R1-O}*10^2$ | 9.191 | 4.303 | 4.532 | 2.864 |
| | | | | |
| R2-O1 x 3 | 2.151(11) | 2.260(7) | 2.160(3) | 2.227(12) |
| R2-O2 x 3 | 2.192(20) | 2.323(11) | 2.182(5) | 2.277(11) |
| R2-O4 | 2.512(69) | 2.459(13) | 2.312(10) | 2.401(10) |
| <R2−O> | 2.220 | 2.315 | 2.191 | 2.273 |
| $\varepsilon_{R2-O}*10^2$ | 12.065 | 6.547 | 5.038 | 5.705 |

* Hexagonal YMnO$_3$ and ScMnO$_3$ generated from the data in Ref. 20(a) Table I and LuMnO$_3$ data obtained from Ref. 20(b). The distortion parameters $\varepsilon_{Mn-O}$, $\varepsilon_{Mn-Mn}$ and $\varepsilon_{R-O}$ are defined as $\varepsilon = [(1/N)\sum(R - <R>)^2]^{1/2}$.

# Figure Captions

**(Color online) Fig. 1**. Images of a typical synthesized crystal (a), a mounted InMnO$_3$ single crystal for XRD measurements (b) and the crystal structure for InMnO$_3$ (c) are shown.

**(Color online) Fig. 2**. (a) Comparison of the Mn K-edge near-edge spectra of RMnO$_3$ (R=Sc, Y, Lu, In) showing the similar structure for the entire series of compounds with InMnO$_3$ (powdered single crystals) as dashed line. (b) Local structure at the Mn sites from the extended x-ray absorption data. Note the reduced amplitude for the In systems for the distant shells (Mn-Mn third shell).

**(Color online) Fig. 3**. Remanent P-E hysteresis loops of a InMnO$_3$ single crystal measured at 1 kHz at room temperature.

**(Color online) Fig. 4**. (a) Heat capacity (C$_p$/T) *vs.* temperature of single crystals revealing a Néel temperature of approximately 118 K. Insert: shows the shoulder with a peak near 42 K, possibly corresponding to spin rotation of Mn ions (in the ab-plane). The error bars are smaller than the symbols. (b) The low-temperature heat capacity of single crystal InMnO$_3$ plotted as C$_p$/T vs T$^2$. The solid line is fit of expression C$_p$/T = $\gamma$+$\beta$T$^2$+ $\beta_5$T$^4$.

**Fig. 1. Yu** *et al.*

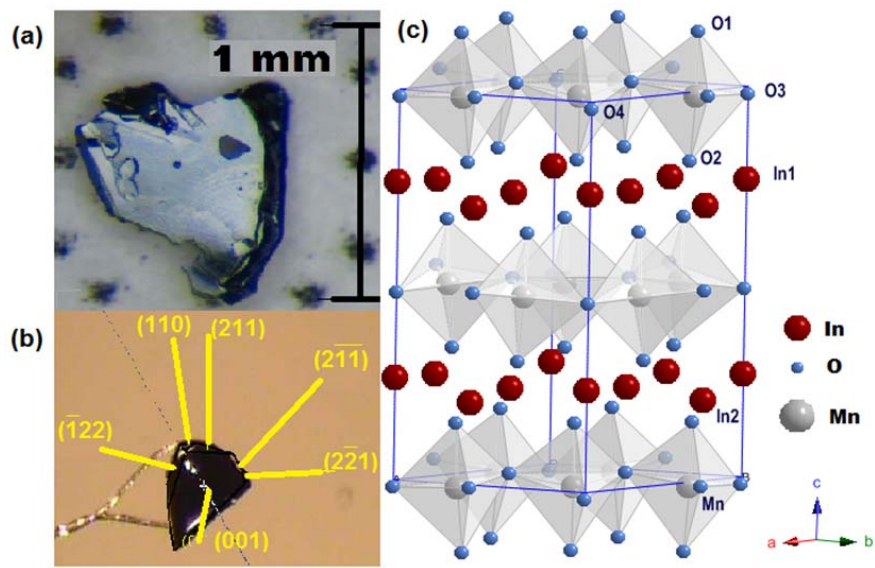

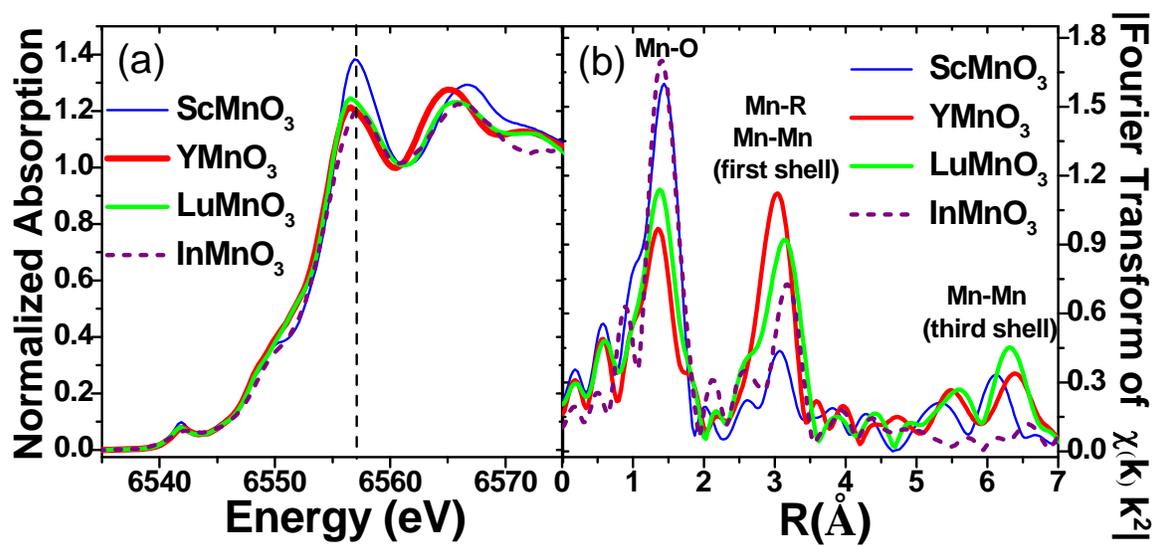

**Fig. 3.** Yu *et al.*

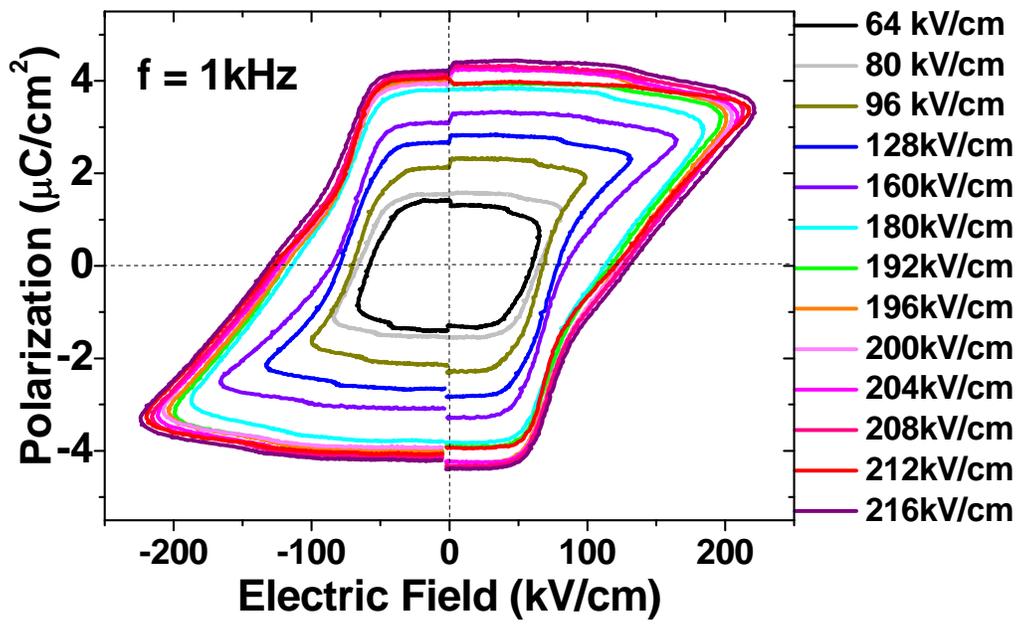

**Fig. 4.** Yu *et al.*

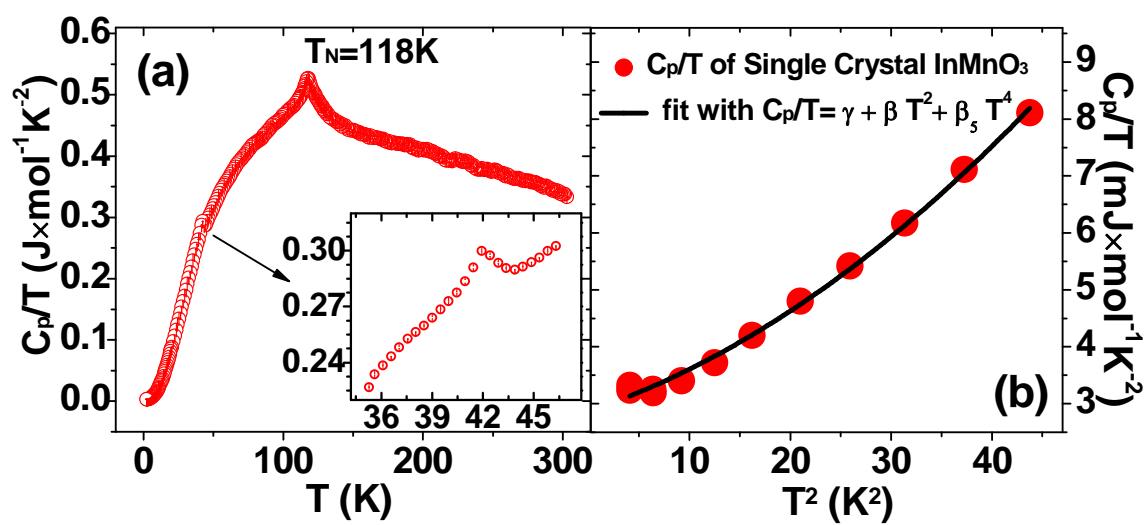